\documentclass[a4paper,cite]{article}
\usepackage{latexsym}

\usepackage[latin1]{inputenc}

\usepackage{graphicx}
\usepackage{color,pst-plot}

\usepackage{amsmath}
\usepackage {amsfonts,amssymb,amstext}
\newcommand{\lbl}[1]{\label{eq:#1}}
\newcommand{ \rf}[1]{(\ref{eq:#1})}

\newcommand{\be}{\begin{equation}}
\newcommand{\ee}{\end{equation}}
\newcommand{\bea}{\begin{eqnarray}}
\newcommand{\eea}{\end{eqnarray}}
\newcommand{\setl}{\setlength\arraycolsep{2pt}}

\newcommand{\noi}{\noindent}
\newcommand{\nn}{\nonumber}
\newcommand{\ra}{\rightarrow}
\newcommand{\Ra}{\Rightarrow}

\newcommand{\cA}{{\cal A}}

\newcommand{\cC}{{\cal C}}

\newcommand{\cF}{{\cal F}}

\newcommand{\cM}{{\cal M}}

\newcommand{\cO}{{\cal O}}

\newcommand{\cR}{{\cal R}}

\newcommand{\Imm}{\mbox{\rm Im}}
\newcommand{\Ree}{\mbox{\rm Re}}

\newcommand{\GeV}{\mbox{\rm GeV}}

\newcommand{\annd}{\mbox{\rm and}}

\newcommand{\als}{\alpha_{\mbox{\rm {\scriptsize s}}}}

\newcommand{\Nc}{\mbox{${\rm N_c}$}}

\input epsf

\usepackage{latexsym}

\usepackage[T1]{fontenc}

\usepackage[latin1]{inputenc}

\usepackage{graphicx}
\usepackage{color,pst-plot}

\usepackage{amsmath}
\usepackage {amsfonts,amssymb,amstext}
\def\theequation{\arabic{section}.\arabic{equation}}

\allowdisplaybreaks[1]
\begin{document}

\begin{titlepage}

\begin{flushright}
\today  \\
\end{flushright}

\vspace*{0.2cm}
\begin{center}

{\LARGE\bf Hadronic Vacuum Polarization in QCD\\[0.5cm]
and its Evaluation in the Euclidean}\\[2cm]

{\large\bf Eduardo de Rafael}\\[1cm]

  {\it Aix-Marseille Universit\'e, CNRS, CPT, UMR 7332, 13288 Marseille, France}
    
\end{center} 

\vspace*{3.0cm}   

\begin{abstract}
We discuss a new technique to evaluate integrals of QCD Green's functions in the Euclidean based on their Mellin-Barnes representation. We present as a first application the evaluation of the lowest order Hadronic Vacuum Polarization (HVP) contribution  to the anomalous magnetic moment of the muon $\frac{1}{2}(g_{\mu}-2)_{\mbox{\rm\tiny HVP}}\equiv a_{\mu}^{\rm HVP}$. 
It is shown that with a precise determination of  the slope and curvature of the HVP function at the origin from lattice QCD (LQCD),  one can already obtain a result for $a_{\mu}^{\rm HVP}$ which may serve as a test of the determinations based on   experimental measurements of  the $e^+ e^-$ annihilation  cross-section into hadrons.
\end{abstract}

\end{titlepage}

%%%%%%%%%%%%%%%%%%%%%%%%%%%%%%%%%%
\section{\Large Introduction}\lbl{int}
\setcounter{equation}{0}
\def\theequation{\arabic{section}.\arabic{equation}}

\noi
In this and forthcoming papers we shall be concerned with QCD two-point functions of  color singlet local operators with possible insertions of soft operators. These Green's functions, weighted by appropriate known functions, when integrated over the full range of their Euclidean momenta dependence,  govern the hadronic contribution to many electromagnetic and weak interaction processes which appear as low energy observables.   
Two well known examples, with no soft insertions, are:

\begin{enumerate}

	\item The Hadronic Vacuum Polarization two-point function:
\be\lbl{eq:EM}
\Pi_{\mu\nu}(q)=i\int d^4 x\  e^{iq\cdot x}
\langle 0\vert T\left(J_{\mu}(x)J_{\nu}
(0)\right)\vert 0\rangle=(q_{\mu}q_{\nu}-q^2 g_{\mu\nu})\Pi(Q^2)\,,\quad Q^2 =-q^2 \ge 0\,,
\ee
where $J_{\mu}(x)$ denotes the hadronic electromagnetic current in the Standard Model. For the light quarks $u$, $d$, $s$,
\be
J_{\mu}(x)=(-i e)\left\{\frac{2}{3}\bar{u}(x)\gamma^{\mu}u(x)-\frac{1}{3}\bar{d}(x)\gamma^{\mu}d(x)-\frac{1}{3}\bar{s}(x)\gamma^{\mu}s(x)\right\}\,.
\ee

\item The Left-Right two-point function in the chiral limit:
\be
\hspace*{-1cm}
\Pi_{\rm LR}^{\mu\nu}(q) =  2i\int d^4 x\,e^{iq\cdot x}\langle 0\vert
 T\left(L^{\mu}(x)R^{\nu}(0)^{\dagger}\right)\vert 0\rangle 
  =  (q^{\mu}q^{\nu}-g^{\mu\nu}q^2)\Pi_{\rm LR}(Q^2)\,,\quad Q^2 =-q^2 \ge 0\,,
\ee
where
\begin{equation} 
L^{\mu}(x)=\bar{d}(x)\gamma^{\mu}\frac{1}{2}(1-\gamma_{5})u(x)
\quad \annd \quad 
R^{\mu}(x)=\bar{d}(x)\gamma^{\mu}\frac{1}{2}(1+\gamma_{5})u(x)\,.
\end{equation} 

\end{enumerate}

\noi
To lowest order in the electromagnetic coupling, the Hadronic Vacuum Polarization (HVP) contribution  to the anomalous magnetic moment of the muon $\frac{1}{2}(g_{\mu}-2)_{\mbox{\rm\tiny HVP}}\equiv a_{\mu}^{\rm HVP}$ is governed by a weighted integral~\cite{LPdeR72,EdeR94} of the hadronic photon self-energy function $\Pi(Q^2)$ in the Euclidean, renormalized on-shell at $Q^2=0$:
\be\lbl{eq:LdeR}
a_{\mu}^{\rm HVP} =  
\frac{\alpha}{\pi}\int_{0}^{1} dx (1-x)
\left[-\Pi\left(Q^2 \equiv \frac{x^2}{1-x}m_{\mu}^2 \right) \right]\,.
\ee
This integral requires knowing the function $\Pi(Q^2)$ all the way from $Q^2 =0$ ($x=0$) to $Q^2 =\infty$ ($x=1$).
In the second example, the $\pi^+ -\pi^0$ mass difference in the Standard Model, to lowest order in the electroweak coupling and in the chiral limit, is governed by a weighted integral of the function $\Pi_{\rm LR}(Q^2)$ in the Euclidean over the full range $0\le Q^2 \le \infty$~(see ref.~\cite{KPdeR98} and references therein):
\be\lbl{eq:ISM1}
(m_{\pi^+}^2-m_{\pi^0}^2)\vert_{\rm SM}=\frac{\alpha}{\pi}\frac{3}{4 f_{\pi}^2}\int_{0}^{\infty} dQ^2\left(1-\frac{Q^2}{Q^2 +M_{Z}^2}\right) \left(-Q^2\Pi_{\rm LR}(Q^2)\right)\,,
\ee
The first term in the r.h.s. produces a mass difference of electromagnetic  origin~\cite{Lowetal67} while the second one corresponds to the small contribution induced by the electroweak $Z$ gauge boson propagator~\cite{KPdeR98}.
Other examples of two-point functions with soft insertions are described e.g. in ref.~\cite{EdeR03} where references to the relevant literature can also be found. 

The purpose of this paper is to describe  a new approach to evaluate Euclidean QCD integrals of this type. 
The problem of computing analytically Green's functions like $\Pi(Q^2)$ and $\Pi_{\rm LR}(Q^2)$  in QCD is due to the fact that QCD perturbation theory (pQCD) is only applicable to short-distances i.e. large $Q^2$ values and, in fact, in the case of Green's functions which are {\it order parameters of spontaneous chiral symmetry breaking}, like  $\Pi_{\rm LR}(Q^2)$, pQCD gives no contribution at all in the chiral limit. The Green's functions in question, however, obey dispersion relations which relate their values in the Euclidean to integrals over spectral functions defined in the Minkowski domain. 
The hadronic photon self-energy $\Pi(Q^2)$ above is not an order parameter of spontaneous chiral symmetry breaking and it  obeys the subtracted dispersion relation
\be\lbl{eq:Pi}
\Pi(Q^2) =  \int_{4 m_{\pi^{\pm}}^2}^{\infty}\frac{dt}{t}\,
\frac{-Q^2}{t+Q^2}\frac{1}{\pi}\Imm\Pi(t)\,, 
\ee
while $\Pi_{\rm LR}(Q^2)$ obeys an unsubtracted dispersion relation
\be
\lbl{eq:Idisplr}
\Pi_{\rm LR}(Q^2) =  \int_{ m_{\pi^{\pm}}^2}^{\infty}dt\frac{1}{t+Q^2}
\frac{1}{\pi}\Imm\Pi_{\rm LR}(t)\,.
\ee

\noi
Furthermore, the two spectral functions $\frac{1}{\pi}\Imm\Pi(t)$ and $\frac{1}{\pi}\Imm\Pi_{\rm LR}(t)$ are accessible to experiment. The first one is directly accessible to experiment via the one photon $e^+ e^-$ annihilation cross section into hadrons ($m_e \ra 0$):
\be\lbl{eq:sigmaee}
\sigma(t)=\frac{4\pi^2 \alpha}{t}\frac{1}{\pi}\Imm\Pi(t)\,,
\ee
and this is in fact the way which $a_{\mu}^{\rm HVP}$ has been evaluated to a high degree of precision~\cite{Davier11, Hagiwara11,Davier16}.
Similarly,  the hadronic spectral function $\frac{1}{\pi}\Imm\Pi_{\rm LR}(t)$, which is the difference:
\be
\frac{1}{\pi}\Imm\Pi_{\rm LR}(t)=\frac{1}{\pi}\Imm\Pi_{\rm VV}(t)-\frac{1}{\pi}\Imm\Pi_{\rm AA}(t)
\ee
of Vector-Vector and Axial-Axial correlation functions, is  accessible via $e^+ e^-$ annihilation and hadronic $\tau$-decay data. QCD perturbation theory contributes both to the VV and to the AA correlation functions, but the contribution is the same in the chiral limit and vanishes in the difference.

The two examples above are, however, rather exceptional. In most cases,  the associated spectral functions to the Green's functions one is interested in are not accessible to experimental determination. In that sense, both $\Pi(Q^2)$ and $\Pi_{\rm LR}(Q^2)$ provide excellent theoretical laboratories to test non perturbative approaches to the determination of the more general type of Green's functions we are concerned with. 

This first paper is dedicated to the study of the hadronic photon self-energy $\Pi(Q^2)$ and its contribution to $a_{\mu}^{\rm HVP}$.
There is a persistent discrepancy between the latest experimental determination of the anomalous magnetic moment of the muon~\cite{BNL}:
\be
a_{\mu}^{\rm exp}=116~592~089(63)\times 10^{-11} \,,
\ee
and its theoretical evaluation in the Standard Model (see e.g. ref~\cite{TH} for a recent description of the various contributions as well as for earlier references.)  
The lowest order HVP contribution to $a_{\mu}^{\rm th}$, evaluated from a  combination of experimental results on 
$e^+ e^-$ data: 
\be\lbl{eq:HVPexps}
a_{\mu}^{\rm HVP}=(6.923\pm 0.042)\times 10^{-8}~{\cite{Davier11}}
\quad \annd \quad 
a_{\mu}^{\rm HVP}=(6.949\pm 0.043)\times 10^{-8}~{\cite{Hagiwara11}}\,,
\ee
gives at present the contribution  with the largest error. The total Standard Model contribution corresponding to these HVP determinations are 
\be
a_{\mu}^{\rm SM}=116~591~802(49)\times 10^{-8}~{\cite{Davier11}} \quad\annd\quad
a_{\mu}^{\rm SM}=116~591~828(50)\times 10^{-8}~{\cite{Hagiwara11}}\,,
\ee
which are significantly lower than the experimental result. A more precise recent determination of $a_{\mu}^{\rm HVP}$:
\be
a_{\mu}^{\rm HVP}=(6.926\pm 0.033)\times 10^{-8}~{\cite{Davier16}}\,,
\ee
confirms this discrepancy.

The possibility of a totally different evaluation of $a_{\mu}^{\rm HVP}$ based on lattice QCD (LQCD, see e.g. refs.~\cite{Burger14,Ch16,Lellouch16} and references therein), may eventually serve as a test of the above  results. This, and the planned experiments at Fermilab and JPARC to measure $a_{\mu}$ in the near future,  which aim at reducing the present uncertainty in $a_{\mu}^{\rm exp}$ by a factor of four,  has prompted a renewed activity on this topic.

This paper is organized as follows. In the next section we discuss the Mellin-Barnes representation of the HVP- function $\Pi(Q^2)$. Section III recalls various equivalent representations which can be used to evaluate $a_{\mu}^{\rm HVP}$. Section IV is dedicated to  an application of Ramanujan's Master Theorem to the HVP-function and Section V to  the Marichev  Interpolation of the Mellin transform of the HVP-Spectral Function, which is in fact the approach that we propose.  Our procedure to evaluate $a_{\mu}^{\rm HVP}$ is discussed in Section VI. First in the case where only the first Mellin  moment $\cM(0)$ is known, i.e. when only the slope of $\Pi(Q^2)$ at the origin is known; and also when both the first two Mellin Moments $\cM(0)$ and $\cM(-1)$ are known, i.e. when the slope and curvature of $\Pi(Q^2)$ at the origin are known.  We illustrate the Marichev Interpolation Approach with the example of Vacuum Polarization in QED, and we test it with a phenomenological {\it Toy Model}~\cite{ToyM} which reproduces the basic features of the hadronic spectral function. We then apply the same interpolation approach to the BHLS model of ref.~\cite{BDDJ16} as well as to a recent LQCD evaluation~\cite{Lellouch16} of the first two moments $\cM(0)$ and $\cM(-1)$, and finally we conclude. 

%%%%%%%%%%%%%%%%%%%%%%%%%%%%%%%%%%
\section{\Large Mellin-Barnes Representation of the HVP-Function.}
\setcounter{equation}{0}
\def\theequation{\arabic{section}.\arabic{equation}}

\noi
We shall extensively use the fact that the electromagnetic  hadronic self-energy function $\Pi(Q^2)$ in Eq.~\rf{eq:Pi} has a useful representation in terms of the Mellin transform of its spectral function $\frac{1}{\pi}\Imm\Pi(t)$ defined as follows~\cite{EdeR14}:
\be\lbl{eq:MPi}
\cM\left[\frac{1}{\pi}\Imm\Pi(t)\right](s)\equiv \cM(s) =\int_{t_0}^\infty\frac{dt}{t}\left(\frac{t}{t_0} \right)^{s-1}\frac{1}{\pi}\Imm\Pi(t)\,,\quad t_0=4 m_{\pi^{\pm}}^2\,,\quad -\infty\le \Ree(s) <1  \,,
\ee
where we have normalized the spectral function $t$-variable to the two-pion threshold value~\footnote{Notice that in ref.~{\cite{EdeR14}} the chosen normalization scale is the muon mass.}. With this normalization $\cM(s)$ is dimensionless and monotonously decreasing in the range: $-\infty\le \Ree(s) <1 $. 

In QCD, the Mellin transform $\cM(s)$ is singular at $s=1$ with a residue which is fixed by pQCD. The contribution from the three light $u$, $d$, $s$ quarks gives
\be\lbl{eq:QCDMs1}
\cM(s)\underset{{s\ra\ 1}}{\thicksim} \left(\frac{\alpha}{\pi}\right)\left(\frac{4}{9}+\frac{1}{9}+\frac{1}{9}\right)N_c\  \frac{1}{3}\left[1+\cO(\als)\right]\ \frac{1}{1-s}\,.
\ee
The representation of $\Pi(Q^2)$ in terms of $\cM(s)$  follows from inserting the Mellin-Barnes  identity
\be
\frac{1}{1+\frac{Q^2}{t}}=\frac{1}{2\pi i}\int\limits_{c_s-i\infty}^{c_s+i\infty}ds\ \left(\frac{Q^2}{t}\right)^{-s}\ \Gamma(s)\Gamma(1-s)
\ee
in the integrand of the r.h.s. of the dispersion relation in Eq.~\rf{eq:Pi}. Then, one has

{\setl
\bea\lbl{eq:PiMB}
\Pi(Q^2) & = &  -Q^2 	\int_{t_0}^\infty\frac{dt}{t^2}
\ \frac{1}{2\pi i}\int\limits_{c_s-i\infty}^{c_s+i\infty}ds\ \left(\frac{Q^2}{t} \right)^{-s} \Gamma(s)\Gamma(1-s)
\ \frac{1}{\pi}\Imm\Pi(t) \nn \\
& = & -\frac{Q^2}{t_0}\ \frac{1}{2\pi i}\int\limits_{c_s-i\infty}^{c_s+i\infty}ds\ \left(\frac{Q^2}{t_0} \right)^{-s} \Gamma(s)\Gamma(1-s)\ \cM(s)\lbl{eq:MBPi}\,,\quad c_s \equiv \Ree (s)\in ]0,1[\,. 
\eea}

\noi
The interest of this integral representation is encoded in the so called {\it converse mapping theorem} of ref.~\cite{FGD95} (see also refs.~\cite{FGdeR05,AGdeR08,FG12} for applications in QED). This theorem relates the singularities in the complex $s$-plane of the integrand, i.e. the singularities of  $\Gamma(s)\Gamma(1-s)\ \cM(s)$ in our case,  to the  asymptotic expansions of  $\Pi(Q^2)$ for $Q^2$ large and for $Q^2$ small. These relations are as follows:

\begin{itemize}
	\item {\sc Expansion for $Q^2\ra\infty$}
	
In the r.h.s. of the {\it fundamental strip} defined by $c_s \equiv \Ree(s)\in ]0,1[$ in Eq.~\rf{eq:MBPi} i.e. for $\Ree(s)\ge 1$,  the most general	singular expansion\footnote{The singular expansion (or singular series) of a meromorphic function is a formal series collecting the singular elements
at all poles of the function (a singular element being the truncated
Laurent's series at $\mathcal{O}(1)$ of the function at a given pole)
and it is conventionally denoted by the symbol $\asymp$~{\cite{FGD95}}.} of the function $\Gamma(s)\Gamma(1-s)\ \cM(s)$ is of the following type:
\be\lbl{eq:singexp}
\Gamma(s)\Gamma(1-s)\ \cM(s)\asymp\sum_{p=1}\sum_{k=0}\frac{	\mathsf{a}_{p,k}}{(s-p)^{k+1}}\, \quad {\rm where~here} \quad (p,k)\in\mathbb{N}\,.
\ee
The corresponding asymptotic behaviour of $\Pi(Q^2)$ for $Q^2$ large ordered in increasing powers of $t_0 /Q^2$ is then:
\begin{equation}\lbl{eq:opeexp}
	\Pi(Q^2)\underset{{Q^2\ra\infty}}{\thicksim}-\frac{Q^2}{t_0}\sum_{p=1}\sum_{k=0} \frac{(-1)^{k+1}}{k!}\ \mathsf{a}_{p,k}\ \left(\frac{t_0}{Q^2}\right)^{p}\ \log^{k}\frac{Q^2}{t_0}\,.
\end{equation}
 From Eq.~\rf{eq:singexp} there follows that the possible  presence of powers of $\log Q^2$ terms in this expansion is correlated to the possible singular behaviour of the Mellin transform $\cM(s)$ at $s=1,2,3,\cdots$. In particular, the leading behaviour of $\Pi(Q^2)$ for $Q^2 \ra\infty$, which is controlled by pQCD,  corresponds to $p=1$ and $k=1$, and 
\begin{equation}\lbl{eq:opelog}
	\Pi(Q^2)\underset{{Q^2\ra\infty}}{\thicksim}- \mathsf{a}_{1,1}\ \log\frac{Q^2}{t_0}+\cdots\,,
\end{equation}
where, from the light $u$, $d$, $s$ quarks contribution,
\be\lbl{eq:a11}
\mathsf{a}_{1,1}=\left(\frac{\alpha}{\pi} \right)\left(\frac{2}{3}\right)\frac{1}{3}N_c \left[1+\cO(\als)\right]\,. 
\ee
In other words, the singular behaviour of the Mellin transform in Eq.~\rf{eq:MPi} at $s=1$ is correlated with the leading $\log Q^2$ behaviour of $\Pi(Q^2)$ when $Q^2 \ra\infty$ and with the fact that the hadronic spectral function  goes asymptotically to a constant:
\be
\frac{1}{\pi}\Imm\Pi(t) \underset{{t\ra\infty}}{\thicksim}
\left(\frac{\alpha}{\pi} \right)\left(\frac{2}{3}\right)\frac{1}{3}N_c \left[1+\cO(\als)\right]\,.
\ee

	\item {\sc Expansion for $Q^2\ra 0$}

In the l.h.s. of the {\it fundamental strip} i.e. for $s\le 0$, the most general form of the singular expansion of the function $\Gamma(s)\Gamma(1-s)\ \cM(s)$ is of the following type: 
\be
\Gamma(s)\Gamma(1-s)\ \cM(s)\asymp\sum_{p=0}\sum_{k=0}\frac{\mathsf{b}_{p,k}}{(s+p)^{k+1}}\,,
\ee
and the corresponding asymptotic behaviour of $\Pi(Q^2)$ ordered in increasing powers of $Q^2 /t_0$ is then:
\begin{equation}
\Pi(Q^2)\underset{{Q^2\ra 0}}{\thicksim}-\frac{Q^2}{t_0}\sum_{p=0}\sum_{k=0}(-1)^k \ \mathsf{b}_{p,k}\ \left(\frac{Q^2}{t_0}\right)^{p}\log^k \frac{Q^2}{t_0}\,.
\end{equation}
In QCD  the Mellin transform in Eq.~\rf{eq:MPi} for $\Ree(s)<1$ is not singular and, therefore, there are no powers of $\log Q^2$ in the expansion of $\Pi(Q^2)$ for $Q^2 \ra 0$. The expansion in this region is a power series:
\be\lbl{eq:exp0}
\Pi(Q^2)\underset{{Q^2\ra 0}}{\thicksim}-\frac{Q^2}{t_0}\sum_{n
=0}
\ \mathsf{b}_{n}\ \left(\frac{Q^2}{t_0}\right)^{n}\,,
\ee
and the coefficients $\mathsf{b}_{n}$ are fixed by the moments $\cM(s)$ at $s=-n$, with $n=0,1,2,3,\cdots$. From the dispersion relation in Eq.~\rf{eq:Idisplr} there follows that these moments  correspond to successive derivatives of the HVP self-energy $\Pi(Q^2)$ at the origin:
\be\lbl{eq:momeucl}
\cM(-n)=\int\limits_{0}^{\infty}\frac{dt}{t}\left(\frac{t_0}{t} \right)^{1+n}
\frac{1}{\pi}\Imm\Pi(t)=
\frac{(-1)^{n+1} }{(n+1)!}( t_0)^{n+1} \left(\frac{\partial^{n+1}}{(\partial Q^2 )^{n+1}}\Pi(Q^2)\right)_{Q^2 =0}\,.
\ee
More precisely
\be
\mathsf{b}_{n}=(-1)^{n}\cM(-n)\,.
\ee
{\it We conclude that the determination of a few terms of the Taylor expansion of $\Pi(Q^2)$ in LQCD, i.e. of a few derivatives of $\Pi(Q^2)$ at $Q^2 =0$, is equivalent to a determination of the Mellin transform of the physical spectral function at a few discrete values $s=-n$ which we call the Mellin Moments.}  

\end{itemize}

%%%%%%%%%%%%%%%%%%%%%%%%%%%%%%%%%%%%%%%%%%%%%%%%%%%%%
\section{\Large The HVP Contribution to $g_{\mu}-2$.}
\setcounter{equation}{0}
\def\theequation{\arabic{section}.\arabic{equation}}

\noi
There are several equivalent representations of $a_{\mu}^{\rm HVP}$ which we next recall.

\begin{itemize}
	\item{\sc The Standard Representation in terms of the Hadronic Spectral Function}~\cite{BM61}

\be\lbl{eq:standard}
a_{\mu}^{\rm HVP}=\frac{\alpha}{\pi}\int_{t_0}^{\infty}
\frac{dt}{t}\int_{0}^{1}dx\frac{x^2(1-x)}{x^2+\frac{t}{m_{\mu}^2}(1-x)}
\frac{1}{\pi}\Imm\Pi(t)\,.
\ee
This is the traditional representation for a determination of $a_{\mu}^{\rm HVP}$ when using experimental data and/or phenomenological models. 
From this representation one can easily see that the integral over the Feynman parameter $x$ is a function of $\frac{m_{\mu}^2}{t}$ which decreases monotonously as $t$ runs from the hadronic threshold $t_0 =4m_{\pi}^2$ to $t=\infty$. The contribution to $a_{\mu}^{\rm HVP}$ is, therefore, dominated by the low-$t$ behaviour of the spectral function. In fact, from the inequality 
\be
\frac{x^2(1-x)}{x^2+\frac{t}{m_{\mu}^2}(1-x)}\le 
x^2 \ \frac{m_{\mu}^2}{t}\,,
\ee
there follows a rigorous  upper bound for $a_{\mu}^{\rm HVP}$~\cite{BdeR69}:
\be\lbl{eq:BdeR69}
a_{\mu}^{\rm HVP}\le\frac{\alpha}{\pi}\frac{1}{3}\int_{t_0}^{\infty}\frac{dt}{t}\frac{m_{\mu}^2}{t}\frac{1}{\pi}\Imm\Pi(t)=
\frac{\alpha}{\pi}\frac{1}{3}m_{\mu}^2\left(-\frac{\partial\Pi(Q^2)}{\partial Q^2}\right)_{Q^2 =0}\,,
\ee
where the equality in the r.h.s. results from the  relation in Eq.~\rf{eq:momeucl} when $n=0$. In other words, 
$a_{\mu}^{\rm HVP}$ is bounded by the {\it  Slope of the  HVP at the Origin} i.e. by the {\it First Mellin Moment $\cM(0)$.}
As emphasized in ref.~\cite{EdeR14}: {\it ``the comparison between the determinations of $\cM(0)$ from LQCD and experimental results should provide an important first test''.}

Using the dispersion relation in Eq.~\rf{eq:Pi} one can rewrite the parametric representation in Eq.~\rf{eq:standard} as follows:

{\setl
\bea
a_{\mu}^{\rm HVP} & = & \frac{\alpha}{\pi}\int_{t_0}^{\infty}
\frac{dt}{t}\int_{0}^{1}dx\frac{x^2(1-x)}{x^2+\frac{t}{m_{\mu}^2}(1-x)}
\frac{1}{\pi}\Imm\Pi(t)\nn \\
 & = & \frac{\alpha}{\pi} \int_{0}^{1}dx (1-x)\int_{t_0}^{\infty}\frac{dt}{t}\frac{\frac{x^2}{1-x}m_{\mu}^2}{t+\frac{x^2}{1-x}m_{\mu}^2 }\frac{1}{\pi}\Imm\Pi(t) \nn \\
 & = &  \frac{\alpha}{\pi}\int_{0}^{1}dx (1-x)\left[-\Pi\left(\frac{x^2}{1-x}m_{\mu}^2 \right)\right]\,,
\eea}

\noi
resulting in the representation for $a_{\mu}^{\rm HVP}$ quoted in Eq.~\rf{eq:LdeR} i.e.

\item{\sc The Representation in terms of the Euclidean Photon Self-Energy}~\cite{LPdeR72,EdeR94}
						
\be\lbl{eq:HVPa}
  a_{\mu}^{\rm HVP}  = 
\frac{\alpha}{\pi}\int_{0}^{1} dx (1-x)
{\left[-\Pi\left(\frac{x^2}{1-x}m_{\mu}^2 \right) \right]\,,\quad Q^2 \equiv\frac{x^2}{1-x}m_{\mu}^2 }\,. \nn
\ee
Trading the Feynman parametric $x$-integration by an integration over the Euclidean $Q^2$-variable results in a more complicated expression 
\be
  a_{\mu}^{\rm HVPL}  = 
\frac{\alpha}{\pi}\int_0^\infty \frac{dQ^2}{Q^2} \sqrt{\frac{Q^2}{4 m_{\mu}^2+Q^2}}\left(\frac{\sqrt{4 m_{\mu}^2 +Q^2}-\sqrt{Q^2}}{{\sqrt{4 m_{\mu}^2 +Q^2}+\sqrt{Q^2}}} \right)^2 [-\Pi(Q^2)]\,,
\ee
which is the one  proposed in ref.~\cite{Blum03} for LQCD determinations of $a_{\mu}^{\rm HVPL}$. This requires, however, an interpolation procedure to evaluate $\Pi(Q^2)$ in the $Q^2$-regions where there is not a direct LQCD evaluation. It is precisely this interpolation procedure which is the main concern of this paper.

\item{\sc The Representation in terms of the Adler Function}

\be\lbl{eq:adg2}
a_{\mu}^{\rm HVP}=\frac{\alpha}{\pi}\ 
\frac{1}{2}\int_{0}^{1} dx\ x (2-x) 
{ \cA\left(Q^2 \equiv\frac{x^2}{1-x}m_{\mu}^2\right)}
\ee
where
\be
\cA (Q^2)=-m_{\mu}^2 \frac{\partial\Pi(Q^2)}{\partial Q^2}\,,
\ee
follows from the one in Eq.~\rf{eq:LdeR} integrating by parts and using the fact that $\Pi(0)=0$.

In terms of the Euclidean $Q^2$-variable:
\be
a_{\mu}^{\rm HVP}=\frac{\alpha}{\pi}\int_0^\infty d\omega\  G(\omega)\ \cA(\omega m_{\mu}^2)\,,
\ee
where
\be
G(\omega)=\frac{1}{4}\left[(2+\omega)(2+\omega-\sqrt{\omega)}\sqrt{4+\omega})-2 \right]\,\quad\annd\quad\omega=\frac{Q^2}{m_{\mu}^2}\,.
\ee
In this representation the upper bound~\cite{BdeR69} in Eq.~\rf{eq:BdeR69} follows from the positivity of the spectral function $\frac{1}{\pi}\Imm\Pi(t)$ which gives rise to the inequality
\be
-\frac{d}{d\omega}\Pi(\left(\omega m_{\mu}^2 \right)\le -\frac{d}{d\omega}\Pi\left(\omega m_{\mu}^2 \right)\vert_{\omega=0}\,,
\ee
and the fact that the  function 
\be
G(\omega)\equiv \frac{1}{4}
\left[\left(2+\omega \right)\left(2+\omega-\sqrt{\omega}\sqrt{4+\omega}\right)-2\right]\,,
\ee
is positive and monotonously decreasing. Therefore

{\setl
\bea
a_{\mu}^{\rm HVP} & \leq &  \frac{\alpha}{\pi}\  \left(-\frac{d}{d\omega}\Pi\left(\omega m_{\mu}^2 \right)\vert_{\omega=0}\right)\int_0^\infty d\omega\frac{1}{4}
\left[\left(2+\omega \right)\left(2+\omega-\sqrt{\omega}\sqrt{4+\omega}\right)-2\right]\nn \\
 & = & \frac{\alpha}{\pi}\ \frac{1}{3} \left(-\frac{d}{d\omega}\Pi\left(\omega m_{\mu}^2 \right)\vert_{\omega=0}\right)=\frac{\alpha}{\pi}\ \frac{1}{3}\int_{4 m_{\pi}^2}^\infty\frac{dt}{t}\frac{m_{\mu}^2}{t}\frac{1}{\pi}\Imm\Pi(t)\,.\lbl{eq:BdeRM}
\eea}

\noi
The function $G(\omega)$ has the following asymptotic behaviours:
\be 
G(\omega)\underset{{\omega\ra 0}}{\thicksim}\frac{1}{2}-\sqrt{\omega}+\omega -\frac{5}{8}\omega^{3/2}+\frac{1}{4}\omega^2 +\cO[\omega^{5/2}]\,,
\ee
and 
\be 
G(\omega)\underset{{\omega\ra \infty}}{\thicksim}\frac{1}{2\omega^2}-\frac{2}{\omega^3}+\frac{7}{\omega^4} -\frac{24}{\omega^5} +\cO[\frac{1}{\omega^{11/2}}]\,.
\ee

\item
{\sc The Mellin-Barnes integral representation}~\cite{EdeR14}

Inserting the Mellin-Barnes expression for $\Pi(Q^2)$that we obtained in Eq.~\rf{eq:PiMB} in  the Euclidean representation of Eq.~\rf{eq:HVPa}, and explicitly integrating over the $x$-parameter, results in a Mellin-Barnes representation  for $a_{\mu}^{\rm HVP}$
\be\lbl{eq:MBexp}
a_{\mu}^{\rm HVP} 
=  \left(\frac{\alpha}{\pi}\right)\frac{m_{\mu}^2}{t_0} \frac{1}{2\pi i}\int\limits_{c_s-i\infty}^{c_s+i\infty}ds\left(\frac{m_{\mu}^2}{t_0} \right)^{-s} \cF(s)\ \cM(s)\,,
\ee
where $\cF(s)$ denotes the product of Gamma-functions
\be\lbl{eq:FFunction}
\cF(s)= -\Gamma(3-2s)\Gamma(-3+s)\Gamma(1+s)\,.
\ee 
Applying the {\it converse mapping theorem} to this representation, results in a  series expansion of $a_{\mu}^{\rm HVP}$ in powers of $\frac{m_{\mu}^2}{t_0}$, with coefficients which are governed by the values of $\cM(s)$ at $s=0,-1,-2,\cdots$ and give positive contributions;  and by the  values of  the first derivative of $\cM(s)$: 
\be
{\tilde{\cM}(s)}=-\frac{d}{ds}\cM(s)=\int_{t_0}^\infty \frac{dt}{t}\left(\frac{t}{t_0} \right)^{s-1}\log\frac{t_0}{t}\ \frac{1}{\pi}\Imm\Pi(t)
\ee
at $s=-1,-2,\cdots$ which give negative contributions: 

{\setl
\bea\lbl{eq:MAn}
a_{\mu}^{\rm HVP} &  = & \left(\frac{\alpha}{\pi}\right) \frac{m_{\mu}^2}{t_0}\left\{ \frac{1}{3}\cM(0) +\frac{m_{\mu}^2}{t_0}\left[\left(\frac{25}{12}-\log\frac{t_0}{m_{\mu}^2}\right) \cM(-1)+
\tilde{\cM}(-1)\right]\right. \nn \\
 & + &  \left(\frac{m_{\mu}^2}{t_0}\right)^2
\left[\left(\frac{97}{10}-\log\frac{t_0}{m_{\mu}^2}\right) \cM(-2)+6
\tilde{\cM}(-2)\right] \nn \\
 & + & \left.\left(\frac{m_{\mu}^2}{t_0}\right)^3
\left[\left(\frac{208}{5}-\log\frac{t_0}{m_{\mu}^2}\right) \cM(-3)+28
\tilde{\cM}(-3)\right] +\cO\left(\frac{m_{\mu}^2}{t_0}\right)^4\right\}\,.
\eea}

\noi
The bulk of the overall contribution to $a_{\mu}^{\rm HVP}$ comes in fact from just the first few terms. The first term is the upper-bound of ref.~\cite{BdeR69} with successive fast improvements from the following terms.
This expansion, when tested with phenomenological models~\cite{ToyM,BDDJ16}, reproduces the full answer to a good accuracy  with just the terms in the first two lines. 

The Mellin Moments $\cM(n)$ and their derivatives ${\tilde{\cM}(n)}$ at $n= 0,-1,-2,\cdots$ provide tests for a comparison between experimental results, phenomenological models and combined LQCD-Pad\'e determinations.  They can also be used as an alternative way to evaluate integrals like the ones we shall encounter later in Eqs.~\rf{eq:mb1mar} and \rf{eq:mb2mar}.

\end{itemize}

%%%%%%%%%%%%%%%%%%%%%%%%%%%%%%%%%%%%%%%%%%%%%%%%%%%%%%%%%%%%
\section{\Large Ramanujan's Master Theorem}
\setcounter{equation}{0}
\def\theequation{\arabic{section}.\arabic{equation}}

\noi
Let us reconsider the Taylor series expansion in Eq.~\rf{eq:exp0} which, as we have shown, is governed by the Mellin Moments $\cM(-n)$, $n=0,1,2,\cdots\,,$ i.e.
\be\lbl{eq:raman}
-\frac{t_0}{Q^2}\Pi(Q^2)\underset{{Q^2\ra 0}}{\thicksim}\left\{
\cM(0)-\frac{Q^2}{t_0}\cM(-1)+\left(\frac{Q^2}{t_0}\right)^2 \cM(-2)-\left(\frac{Q^2}{t_0}\right)^3 \cM(-3)+\cdots \right\}\,.
\ee
This expansion provides the basis for an application of {\it Ramanujan's Master Theorem}~\cite{Raman} to our case. It  states that:

{\setl
\bea\lbl{eq:RTHVP}
& & \int_0^\infty d\left(\frac{Q^2}{t_0}\right)\left(\frac{Q^2}{t_0}\right)^{s-1}\left\{
\cM(0)-\frac{Q^2}{t_0}\cM(-1)+\left(\frac{Q^2}{t_0}\right)^2 \cM(-2)+\left(\frac{Q^2}{t_0}\right)^3\cM(-3)+\cdots \right\}\nn \\
& &\nn \\ 
& & =\ \Gamma(s)\Gamma(1-s)\cM(s)\,.
\eea}

\noi 
This integral identity, which follows from the inverse transform in Eq.~\rf{eq:PiMB}, 
is at the basis of the approach that we are going to use. 

For pedagogical purposes let us first apply it to the simple case of Vacuum Polarization in QED. 

\vspace*{0.25cm}
\subsection {\large Application to Vacuum Polarization in QED}
\vspace*{0.25cm}

Consider the Euclidean behaviour of vacuum polarization in QED for a fermion of mass $m$ which, to lowest order in $\alpha$, is given by the simple Feynman parametric integral
\be
\Pi^{\rm QED}(Q^2)
  =  -\frac{\alpha}{2\pi}\int_0^1 dy (1-y^2)\log\left[1+\frac{Q^2}{4m^2}(1-y^2) \right]\,.
\ee
For $Q^2$-small it has the Taylor series expansion
\be
	-\frac{4m^2}{Q^2}\Pi^{\rm QED}(Q^2)\underset{{Q^2\ra 0}}{\thicksim}\frac{\alpha}{2\pi} \sum_{n=0} \left(\frac{Q^2}{4m^2}\right)^n \frac{(-1)^{n}}{n+1}  \int_0^1 dy (1-y^2)^{2+n}\,,
\ee
and since
\be
\int_0^1 dy (1-y^2)^{2+n}=\frac{\sqrt{\pi}}{2}\frac{\Gamma(3+n)}{\Gamma(\frac{7}{2}+n)}\,,
\ee
it can be expressed as follows
\be\lbl{eq:taylorqed}
	-\frac{4m^2}{Q^2}\Pi^{\rm QED}(Q^2)\underset{{Q^2\ra 0}}{\thicksim} \sum_{n=0} (-1)^n  \left(\frac{Q^2}{4m^2}\right)^n \left\{\frac{\alpha}{2\pi}\frac{1}{n+1}\frac{\sqrt{\pi}}{2}\frac{\Gamma(3+n)}{\Gamma(\frac{7}{2}+n)}\right\}\,.
\ee
Recall that, as discussed earlier, the successive derivatives of $\Pi^{\rm QED}(Q^2)$ at $Q^2 \ra 0$ are given by the Mellin Moments $\cM^{\rm QED}(-n)$, $n=0,1,2,\cdots$, of the QED spectral function. 

The application of Ramanujan's Master Theorem to this Taylor series  is straightforward. First, it states that the Mellin transform of $-\frac{4m^2}{Q^2}\Pi^{\rm QED}(Q^2)$ is related to the {\it full} Mellin transform of the spectral function $\cM^{\rm QED}(s)$ as follows:
\be
\int_0^\infty d\left(\frac{Q^2}{4m^2}\right)\left(\frac{Q^2}{4m^2}\right)^{s-1}\left(	-\frac{4m^2}{Q^2}\Pi^{\rm QED}(Q^2)\right)=\Gamma(s)\Gamma(1-s)\cM^{\rm QED}(s)\,,
\ee
and, {\it furthermore},  that the {\it full} Mellin transform function $\cM^{\rm QED}(s)$  can be {\it simply}  obtained by the replacement $n\ra -s$ in the $n$-dependent coefficient of the previous Taylor series.
By simple inspection of the Taylor series we conclude, without having to do any integral,  that
\be\lbl{eq:mellinqed}
\cM^{\rm QED}(s)\equiv\int_{4m^2}^\infty \frac{dt}{t}\left(\frac{t}{4m^2}\right)^{s-1}\frac{1}{\pi}\Imm\Pi^{\rm QED}(t)=\frac{\alpha}{\pi}\frac{1}{3}\frac{1}{1-s}\frac{3\sqrt{\pi}}{4}\frac{\Gamma(3-s)}{\Gamma(\frac{7}{2}-s)}\,,
\ee
where $\frac{1}{\pi}\Imm\Pi^{\rm QED}(t)$ is the lowest order QED spectral function
\be
\frac{1}{\pi}\Imm\Pi^{\rm QED}(t)=\frac{\alpha}{\pi}\frac{1}{3}\left(1+\frac{2m^2}{t}\right)\sqrt{1-\frac{4 m^2}{t}}\theta(t-4m^2)\,.
\ee

The function $\cM^{\rm QED}(s)$ thus obtained is the analytic continuation to the complex $s$-plane of the function defined in the region $s<1$ (i.e. the {\it fundamental strip}) where the Mellin Moments $\cM^{\rm QED}(-n)$, $n=0,1,2,\cdots$ are well defined by direct integration of  the QED spectral function.

We shall come back to this simple example later on.

\vspace*{0.25cm}
\subsection {\large Ramanujan's Theorem and the HVP-Function}
\vspace*{0.25cm}

Ramanujan's theorem applied to the HVP self-energy function in Eq.~\rf{eq:RTHVP}  guarantees that the incorporation of more and more moments $\cM(-n)$ at integer $n$-values $n=0,1,2,\cdots$  converges  to the full Mellin function $\cM(s)$. 
The fact that these moments are numerically accessible to  LQCD  (at least for the  low $n$-values~\cite{Lellouch16}) via the determination of the derivatives of the Euclidean hadronic self-energy function $\Pi(Q^2)$ at $Q^2 =0$, provides an interesting starting point towards an alternative evaluation of the HVP contribution to $a_{\mu}^{\rm HVP}$ from first principles.

Ramanujan's Theorem, however,   does not tell us which is the best {\it Interpolating Function} we should  use to approximate the exact $\cM(s)$ function when one only knows numerically  a few  $\cM(-n)$ moments. Pad\'e Approximants to $\Pi(Q^2)$~\cite{Perisetal12}, or the method of conformal polynomials~\cite{GMP14}, cannot be the answer because they fail to reproduce the pQCD behaviour at $s=1$ of $\cM(s)$. Pad\'e Approximants to $\Pi(Q^2)$ in the low-$Q^2$-region, up to a ``reasonable'' $Q_{0}^2$-value from which onwards the pQCD prediction for $\Pi(Q^2)$ takes over (see e.g. ref.~\cite{ABCGPT16} and references therein), is a possible way to proceed but to our knowledge it has not been proved to be the best interpolation procedure. 

These considerations have prompted us to investigate  alternative approaches based on functional interpolations of Mellin Moments which respect known properties of QCD, in particular the fact that $\cM(s)$ is singular at $s=1$. In the next section we present a new technique in this direction inspired by Marichev's class of Mellin transforms~\cite{Mari82} (see also ref.~\cite{Poul10}, Chapter 12),  shown to be applicable to a large class of functions. 
Although we cannot prove that this approach is the best interpolation procedure for the  QCD Green's functions  that we are concerned with, it turns out to be surprisingly successful when tested with the previous QED example, as well as  with  phenomenological models of the hadronic spectral function which we later discuss.

%%%%%%%%%%%%%%%%%%%%%%%%%%%%%%%%%%%%%%%%%%%%
\section{\Large Marichev's Interpolating Approach.}
\setcounter{equation}{0}
\def\theequation{\arabic{section}.\arabic{equation}}

\noi
The most general form of a Mellin transform of Marichev's class is a fraction involving products of Gamma-functions:
\be\lbl{eq:marichev}
\cM(s)=C\ \displaystyle\prod_{i,j,k,l}\frac{\Gamma(a_{i}-s)\Gamma(c_{j}+s)}{\Gamma(b_{k}-s)\Gamma(d_{l}+s)}\,,
\ee
where $C$ and $a_{i}$, $b_{k}$, $c_{j}$ and $d_{l}$ are real constants and the variable $s$ appears only with a $\pm$ coefficient. In our case, these constants will be  adjusted so as to reproduce as well as possible properties that we know of the  QCD Mellin transform of the physical spectral function in Eq.~\rf{eq:MPi}. In particular, the choice of the $\pm$ signs in the $\Gamma$-functions of the interpolating expressions that we shall consider must  result in a  function monotonously decreasing in the range $1>\Ree(s)\ge -\infty$. This excludes interpolations which produce poles and/or zeros in this region. 

The inverse Mellin transform of a function of the Marichev class is a generalized hypergeometric function which can be reconstructed using the Slater procedure~\cite{Slater66}. This way, one can obtain the underlying  spectral function, as well as the underlying self-energy function in the Euclidean, corresponding to a given Marichev interpolation~\footnote{The Slater procedure applied to several examples, as well as the convergence of the Marichev interpolation,  will be discussed in a forthcoming paper~\cite{GdeR17}.}. The fact that practically all the known functions in Mathematical Physics can be expressed as generalized hypergeometric functions gives a strong support to the interpolation approach  that we are advocating.

Let us first  illustrate how the Marichev interpolating approach works  in the simple case of vacuum polarization in QED that we discussed before and where the same requirement of monotonously decreasing for the Mellin transform also applies: 

\vspace*{0.25cm}
\subsection {\large Application to Vacuum Polarization in QED}
\vspace*{0.25cm}

\noi
The QED Mellin transform in Eq.~\rf{eq:mellinqed} is indeed of the Marichev class. We shall show below that,
in this case, the interpolation method we propose converges very fast to the exact result.

\begin{itemize}
	\item Assume that we only know the asymptotic behaviour of the QED spectral function i.e. 
\be
\frac{1}{\pi}\Imm\Pi^{\rm QED}(t)\underset{{t\ra\infty}}{\thicksim}\frac{\alpha}{\pi}\frac{1}{3}\,.
\ee
As already discussed, this implies  that $\cM(s)$ has a pole at $s=1$ with $\frac{\alpha}{\pi}\frac{1}{3}$ as the residue at the pole and, therefore,  fixes what we shall call in this case the   {\it First Marichev Interpolation} to a simple ratio of Gamma-functions: 
\be
\cM^{\rm QED}(s)\Ra\cM^{(1)}(s)= \frac{\alpha}{\pi}\frac{1}{3}\frac{1}{1-s}= \frac{\alpha}{\pi}\frac{1}{3}\frac{\Gamma(1-s)}{\Gamma(2-s)}\,,
\ee
the upper-script $(1)$ in $\cM^{(1)}(s)$ meaning that we have only  used as information the value of $\cM^{\rm QED}(s)$ at $s=1$.

\item The next step will use the information that, besides the singular behaviour at $s=1$  we also know the slope of $\Pi(Q^2)$ at $Q^2 =0$ which, as previously discussed,  is equivalent to say that we know the first Mellin Moment at $s=0$. In QED:
\be
\cM^{\rm QED}(0)=\frac{\alpha}{\pi}\frac{1}{3}\frac{4}{5}\,,
\ee
and with this information we can now improve our ansatz to a   {\it Second Marichev Interpolation}:
\be
\cM^{\rm QED}(s)\Ra\cM^{(1)}_{(0)}(s)=  \frac{\alpha}{\pi}\frac{1}{3}\frac{1}{1-s}\frac{\Gamma(b-1)}{\Gamma(b-s)}
\ee
which, at $s=1$, does not change the singular behaviour of $\cM^{\rm QED}(s)$ and it satisfies the requirement of being a  function monotonously decreasing in the range $1>\Ree(s)\ge -\infty$. The new parameter $b$ can then be fixed from the identity 
\be
\cM^{\rm QED }(0)\Vert^{(1)}_{(0)}=\cM^{\rm QED}(0)\quad \Ra \quad b=\frac{9}{4} \,,
\ee
and therefore
\be\lbl{eq:10MQED}
\cM^{\rm QED}(s)\Vert^{(1),(2)}=\frac{\alpha}{\pi}\frac{1}{3}\frac{1}{1-s}\frac{\Gamma(5/4)}{\Gamma(9/4-s)}\,.
\ee

\item A further improvement results when we add the information that we also know the term of $\cO\left(\frac{M^2}{Q^2} \right)$ in the asymptotic expansion of $\Pi^{\rm QED}(Q^2)$ when $Q^2 \ra\infty$. According to our discussion in Section II this is equivalent to say that 
\be
\cM^{\rm QED}(2)=\frac{\alpha}{\pi}\frac{1}{3}\left(-\frac{3}{2}\right)\,,
\ee
and it allows us to consider an improved  {\it Third Marichev Interpolation} function satisfying the requirement of being monotonously decreasing in the range $1>\Ree(s)\ge -\infty$:
\be\lbl{eq:3MQED}
\cM^{\rm QED}(s)\Ra \cM^{\rm QED}(s)\Vert^{(1),(2)}_{(0)}= \frac{\alpha}{\pi}\frac{1}{3}\frac{1}{1-s}\frac{\Gamma(c-1)}{\Gamma(c-s)}\frac{\Gamma(d-s)}{\Gamma(d-1)}\,.
\ee
The new parameters $c$  and $d$ are fixed by matching this ansatz to the physical values of $\cM^{\rm QED}(s)$ at $s=0$ and $s=2$, which implies the equations:
\be
\frac{d-1}{c-1}=\frac{4}{5}\,,\quad\annd\quad\frac{c-2}{d-2}=\frac{3}{2}\,, 
\ee
with the results
\be
c=\frac{7}{2}\,,\quad\annd\quad d=3\,,
\ee
and therefore
\be
\cM^{(1),(2)}_{(0)}(s)=\frac{\alpha}{\pi}\frac{1}{3}\frac{1}{1-s}\Gamma(5/2)\frac{\Gamma(3-s)}{\Gamma(7/2 -s)}\,.
\ee
Quite remarkably, we find that this {\it Third Marichev Interpolation} already coincides with the {\it Exact Result} for $\cM^{\rm QED}(s)$ in Eq.~\rf{eq:mellinqed}! (recall that $\Gamma(5/2)=\frac{3\sqrt{\pi}}{4}$). If we try to improve the {\it Third Marichev Interpolation} with further information, e.g. the knowledge of $\cM^{\rm QED}(-1)$, we find that the new input value coincides exactly with the one predicted by $\cM^{(1),(2)}_{(0)}(s)$ at $s=-1$ and, therefore, there is no room for further improvement.

\end{itemize}

We have found  that the Marichev interpolation approach that we are advocating,  when applied to vacuum polarization in QED, reproduces the {\it exact expression} of the Mellin transform of the spectral function with just the information provided by  the values of {\it three} Mellin Moments. 	
Encouraged by this remarkable success we propose to apply the same approach to vacuum polarization in QCD which will be discussed in the next subsection; but, before we leave this QED example, we still want to comment on another issue: the calculation of the QED vacuum polarization  contribution to the anomalous magnetic moment of an external fermion. We explain this in the following sub-subsection.

\vspace*{0.25cm}
\subsubsection {From the Mellin transform of the QED Spectral Function\\ to the Anomalous Magnetic Moment}
\vspace*{0.25cm}

For simplicity we shall consider the case where the external fermion is the same as the one which induces the  vacuum polarization contribution. The Mellin-Barnes representation in Eq.~\rf{eq:MBexp} when adapted to this case is as follows    
\be\lbl{eq:MBexp}
a^{\rm QED}({\rm VP}) 
=  \left(\frac{\alpha}{\pi}\right)\frac{1}{4} \frac{1}{2\pi i}\int\limits_{c_s-i\infty}^{c_s+i\infty}ds\left(\frac{1}{4} \right)^{-s} \cF(s)\ \cM^{\rm QED}(s)\,,
\ee
with $\cM^{\rm QED}(s)$ given in Eq.~\rf{eq:mellinqed} and $\cF(s)$ the same function as in Eq.~\rf{eq:FFunction}. Since
$\cM^{\rm QED}(s)$ is explicitly known we can make a direct evaluation of this integral, provided we choose a value for $c_s$ within the {\it fundamental strip}~\footnote{I am very grateful to Santi Peris for reminding me of this fact.} i.e. $c_s \equiv \Ree(s)\in ]0,1[$.
%%%%%%%%%%%%%%%%%%%%%%%%%%%%%
\begin{figure}[!ht]
\begin{center}
\hspace*{-1cm}\includegraphics[width=0.60\textwidth]{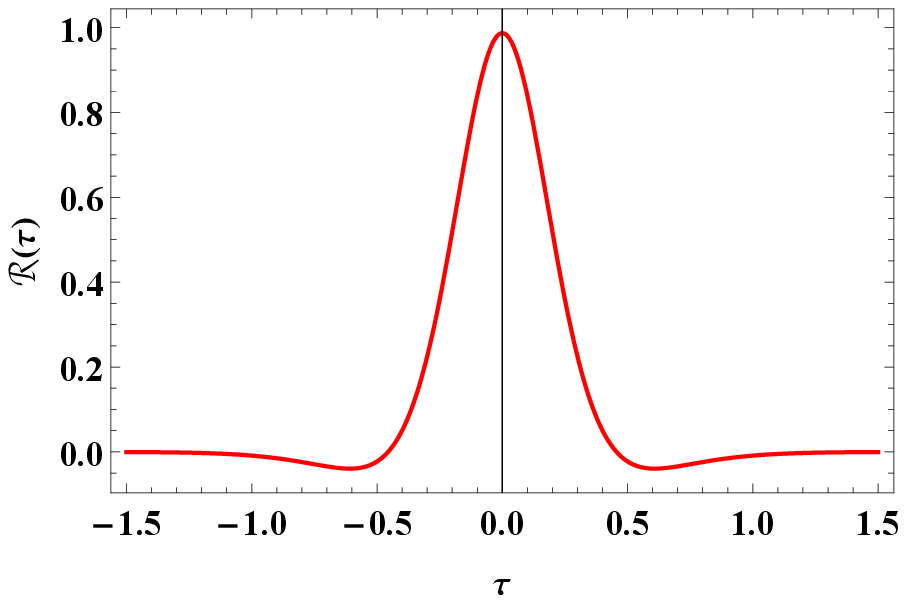} 
\bf\caption{\lbl{fig:tauqed}}
\vspace*{0.25cm}
{\it Shape of the Function $\cR(\tau)$ in the integrand of  Eq.~\rf{eq:qedtau} (first line).}
\end{center}
\end{figure}
%%%%%%%%%%%%%%%%%%%%%%%%%%%%

\noi
With
$
s=\frac{1}{2}-i\tau
$ as a choice,  
and after some simplifications, the previous integral becomes:

{\setl
\bea\lbl{eq:qedtau}
\lefteqn{\hspace*{-2cm} a^{\rm QED}({\rm VP}) = 
 \left(\frac{\alpha}{\pi}\right)^2
\frac{1}{4}\frac{1}{2\pi}\int_{-\infty}^{+\infty} d\tau  \left\{\underbrace{\left[\frac{1}{1+\tau^2}-\frac{10}{4+\tau^2}+\frac{40}{25+\tau^2}\right]\frac{\pi^2}{[\cosh(\pi\tau)]^2}}_{\Re(\tau)} \right. } \nn  \\
& & \left. -i \tau\left[\frac{1}{1+\tau^2}-\frac{5}{4+\tau^2}+\frac{16}{25+4\tau^2} \right] \frac{\pi^2}{[\cosh(\pi\tau)]^2}\right\}\nn \\
& & = \left(\frac{\alpha}{\pi}\right)^2 \times 0.01568742185910\,,
\eea}

\noi
which, to this remarkable accuracy,  agrees with the exact analytic result (see e.g. ref.~\cite{AGdeR08}):
\be
a^{\rm QED}({\rm VP})  =  \left(\frac{\alpha}{\pi}\right)^2
\left(\frac{119}{36}-\frac{\pi^2}{3} \right)\,.
\ee
Notice that the imaginary part of the integrand in the second line of  Eq.~\rf{eq:qedtau} gives zero contribution to the integral. The shape of the function $\cR(\tau)$ in the first line, which is symmetric under $\tau\ra -\tau$, defines the real part of the integrand and it is shown in Fig~\rf{fig:tauqed}.

\vspace*{0.25cm}

\subsection {\large Application to Vacuum Polarization in QCD}
\vspace*{0.25cm}

We can now go back to QCD. 
As already stated in Eq.~\rf{eq:QCDMs1} the QCD Mellin transform $\cM(s)$ is singular at $s=1$ with a residue which is fixed by pQCD. 
For three light $u$, $d$, $s$ quarks and neglecting $\cO(\als)$ corrections this fixes the  pQCD-{\it Marichev Interpolation} to the simple function 
\be
\cM(s)\Ra\cM^{(1)}(s)=  \cC\ \frac{1}{1-s}\,,
\ee
where $\cC$ will denote the overall constant
\be\lbl{eq:constantc}
\cC\equiv \frac{\alpha}{\pi}\frac{1}{3}\ N_c\left(\frac{2}{3}\right)\,.
\ee
Contrary to the QED case discussed before,  the  $\cO(1/Q^2)$ term in the asymptotic expansion of $\Pi(Q^2)$ for large $Q^2$ in QCD  vanishes  in the chiral limit which requires that
\be
\cM(2)=0\,,
\ee
and the corresponding interpolation is then
\be\lbl{eq:12MQCD}
\cM^{(1,2)}(s)=\cC\ \frac{1}{(1-s)\Gamma(2-s)}\,.
\ee
This is as much as we shall use from the short-distance behaviour of QCD.

\vspace*{0.25cm}
\subsubsection{First Marichev Interpolation when $\cM(0)$ is known} 
\vspace*{0.25cm}

Let us now construct the QCD equivalent of the  {\it Marichev Interpolation} which in the QED example above already reproduced the exact result. This corresponds to the case where, besides $\cM(1)$ and $\cM(2)$, we also know the slope at the origin of the HVP-function, i.e. we know $\cM(0)$. The corresponding interpolation, which in the QCD case we shall call the  {\it First Marichev Interpolation} has the same functional form as the QED one in Eq.~\rf{eq:3MQED}, i.e.
\be
\cM(s)\Ra \cM^{(1,2)}_{(0)}(s)= \cC\ \frac{1}{1-s}\frac{\Gamma(c-1)}{\Gamma(c-s)}\frac{\Gamma(d-s)}{\Gamma(d-1)}\,,
\ee
but with the parameters $c$ and $d$ restricted now to satisfy the two QCD constraints at $s=2$ and $s=0$:
\be
\frac{c-2}{d-2}=0\,,\quad\annd\quad \cC\ \frac{d-1}{c-1}=\cM(0)\,,
\ee
which results in
\be
c=2\quad\annd\quad d=1+ \frac{1}{\cC}\ \cM(0)\,.
\ee
With $\cA$ denoting the quantity:
\be\lbl{eq:A}
\cA\equiv \frac{1}{\cC}\cM(0)\,,
\ee
the interpolation in question is then:
\be\lbl{eq:M120}
\cM^{(1,2)}_{(0)}(s)= \cC\ \frac{1}{(1-s)\Gamma(2-s)}\frac{\Gamma\left[ 1+ \cA\ -s\right]}{\Gamma\left(\cA\right)}\,.
\ee

One can now proceed to the determination  of the corresponding prediction for $a_{\mu}^{\rm HVP}$ inserting this  $\cM^{(1),(2)}_{(0)}(s)$ interpolating Mellin transform in Eq.~\rf{eq:MBexp} and evaluating numerically the integral with e.g. the choice $c_s =\frac{1}{2}$:
\be\lbl{eq:mb1mar}
a_{\mu}^{\rm HVP} ({\rm first})
=  \left(\frac{\alpha}{\pi}\right)\frac{m_{\mu}^2}{t_0} \frac{1}{2\pi}\int\limits_{-\infty}^{+\infty}d\tau\left(\frac{m_{\mu}^2}{t_0} \right)^{-\left(\frac{1}{2}-i\tau \right)} \cF\left(\frac{1}{2}-i\tau \right)\ \cM^{(1,2)}_{(0)}\left(\frac{1}{2}-i\tau \right)\,,
\ee
where $\cF(s)$ is the function defined in 
Eq.~\rf{eq:FFunction}.

\vspace*{0.25cm}
\subsubsection {Second Marichev Interpolation when $\cM(0)$ and $\cM(-1)$ are known}
\vspace*{0.25cm}

The previous interpolation can be improved once we know both the slope and the curvature of $\Pi^{\rm HVP}(Q^2)$ at the origin i.e. when $\cM(0)$ and $\cM(-1)$ are known, in which case we use as an ansatz the following {\it Second Marichev Interpolation}:
\be\lbl{eq:M120m1}
\cM(s)\Ra \cM^{(1,2)}_{(0,-1)}(s)= \cC\ \frac{1}{1-s}\frac{1}{\Gamma(2-s)}\frac{\Gamma(e-s)}{\Gamma(e-1)}\frac{\Gamma(f-1)}{\Gamma(f-s)}\,,
\ee
with the $e$ and $f$ parameters fixed by the matching equations:
\be
\cM^{(1,2)}_{(0,-1)}(0)=\cM(0)\quad\annd\quad
\cM^{(1,2)}_{(0,-1)}(-1)=\cM(-1)\,.
\ee
In terms of the quantities $\cA$ (defined in Eq.~\rf{eq:A})  and the ratio:
\be\lbl{eq:R}
\cR=4\frac{\cM(-1)}{\cM(0)}\,,
\ee
we find
\be
f=\frac{1-\cA}{\cR-\cA}\,,\quad\annd\quad e=\cR f\,.
\ee

The corresponding prediction for $a_{\mu}^{\rm HVP}$ is then given by the numerical evaluation of the integral (which is in fact a Fourier-like transform):
\be\lbl{eq:mb2mar}
a_{\mu}^{\rm HVP} ({\rm second})
=  \left(\frac{\alpha}{\pi}\right)\frac{m_{\mu}^2}{t_0} \frac{1}{2\pi}\int\limits_{-\infty}^{+\infty}d\tau\left(\frac{m_{\mu}^2}{t_0} \right)^{-\left(\frac{1}{2}-i\tau \right)} \cF\left(\frac{1}{2}-i\tau \right)\ \cM^{(1,2)}_{(0,-1)}\left(\frac{1}{2}-i\tau \right)\,.
\ee

We wish to emphasize that the first and second Marichev functions in Eqs.~\rf{eq:M120} and \rf{eq:M120m1} are unique in the sense that, with the information provided, they are the most general Mellin transforms satisfying the criteria stated after Eq.~\rf{eq:marichev}.

We have now all the ingredients to test these {\it First} and {\it Second Marichev Interpolations} with phenomenological models, and then to apply them to the evaluation of $a_{\mu}^{\rm HVP}$ using as an input  the recent LQCD determinations~\cite{Lellouch16} of $\cM(0)$ and $\cM(-1)$.

%%%%%%%%%%%%%%%%%%%%%%%%%%%%%%%%%%%%%%%%%%%%%%%%%%%%%%%
\section{\Large Tests with a Phenomenological Model}
\setcounter{equation}{0}
\def\theequation{\arabic{section}.\arabic{equation}}

\noi
In order to test the interpolation approach proposed above, we shall apply it to a phenomenological model of the hadronic spectral function, which we call the {\it Toy Model}~\cite{ToyM}. The model has been constructed to simulate the basic features of the phenomenological spectral function, but with fixed parameters (i.e. no errors) so as to be handled mathematically as an exact function. The {\it Toy Model} spectral function in the low energy range  below $1~\GeV^2$ is shown in Fig.~\rf{fig:toylow}. Although this  {\it Toy Model} should not be confused with the experimental  determination of the spectral function, beautifully shown e.g. in ref.~\cite{Davier16},  it reproduces nevertheless rather well its phenomenological features and, for our purposes, it can be considered as a good theoretical test laboratory. 

%%%%%%%%%%%%%%%%%%%%%%%%%%%%
\begin{figure}[!ht]
\begin{center}
\hspace*{-1cm}\includegraphics[width=0.60\textwidth]{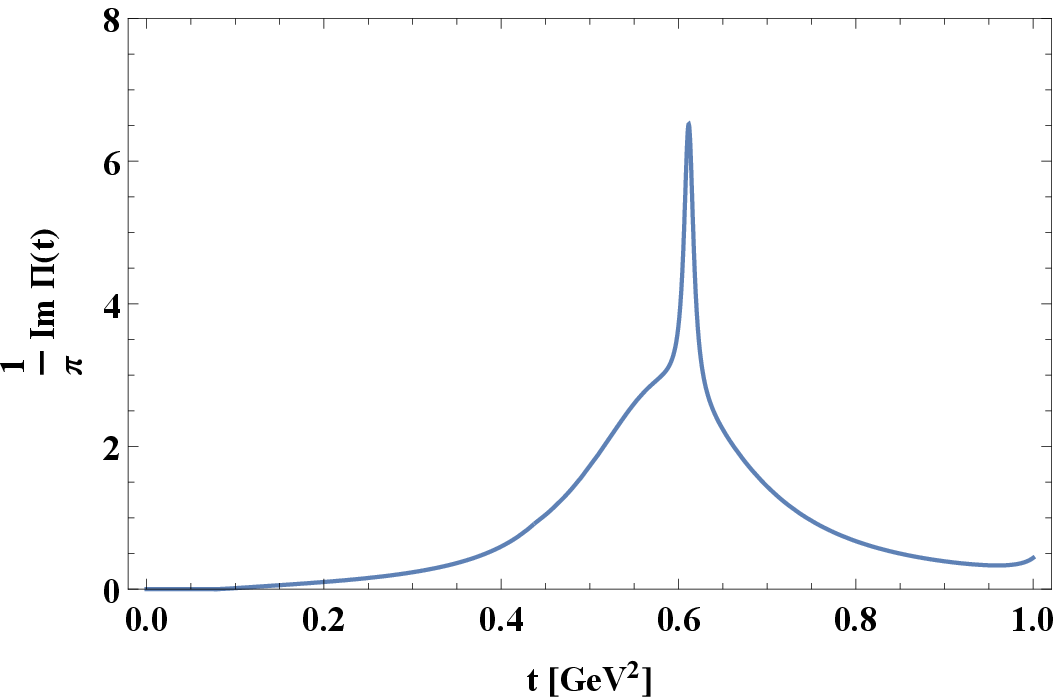} 
\bf\caption{\lbl{fig:toylow}}
\vspace*{0.25cm}
{\it The Toy Model Spectral Function below $1~\GeV^2$}
\end{center}
\end{figure}
%%%%%%%%%%%%%%%%%%%%%%%%%%%%% 
 
We first observe that, by contrast to the complex structure of the spectral function of the {\it Toy Model} shown  in Fig.~\rf{fig:toylow}, its Mellin transform, which is shown in Fig.~\rf{fig:mellintm}, has an extraordinarily smooth shape. In particular, the value at $s=0$, which corresponds to the slope of the $\Pi^{\rm HVP} (Q^2)$-function at the origin and which we shall use later is
\be\lbl{eq:mel0}
\cM^{\rm ToyM}(0)=0.7057904\times 10^{-3}\,.
\ee
The value at $s=-1$, which corresponds to the curvature of the same $\Pi^{\rm HVP} (Q^2)$-function at the origin (the second derivative) and which we shall also use later is
\be\lbl{eq:melm1}
\cM^{\rm ToyM}(-1)=0.1151594\times 10^{-3}\,.
\ee 
For $0\le \Ree(s)<1$ the Mellin transform continues to rise monotonously to become singular at $s=1$ as predicted by pQCD.

%%%%%%%%%%%%%%%%%%%%%%%%%%%%%
\begin{figure}[!ht]
\begin{center}
\hspace*{-1cm}\includegraphics[width=0.60\textwidth]{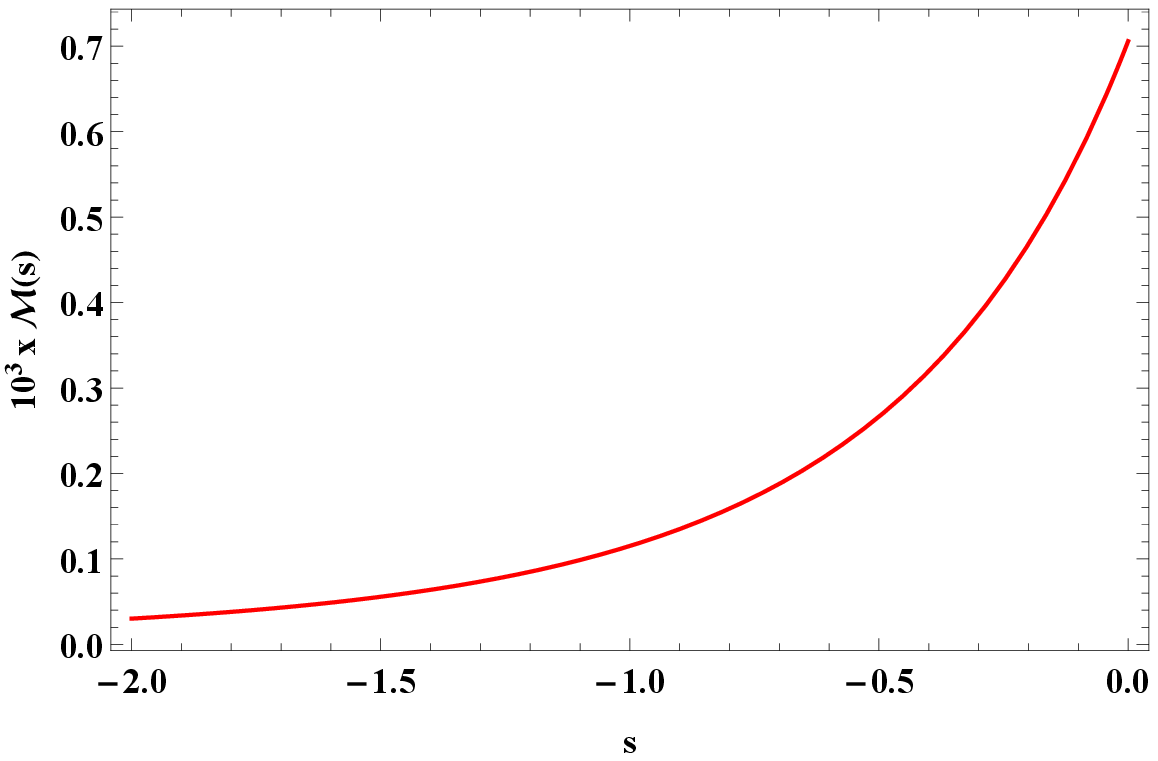} 
\bf\caption{\lbl{fig:mellintm}}
\vspace*{0.25cm}
{\it Mellin Transform of the Toy Model Spectral Function (including charm). }
\end{center}
\end{figure}
%%%%%%%%%%%%%%%%%%%%%%%%%%%%%

Using the standard representation of $a_{\mu}^{\rm HVP}$ in Eq.~\rf{eq:standard} we find that the {\it Toy Model} predicted value of the HVP contribution  to the muon anomaly is
\be\lbl{eq:amutoy}
a_{\mu}^{\rm HVP}({\rm ToyM})=6.812175\times 10^{-8}\,,
\ee
somewhat smaller than the determinations using $e^+ e^-$ data in Eqs.~\rf{eq:HVPexps} which, in fact, have some extra contributions included; but,  as already said, the purpose of the model is not to reproduce experimental results but rather to be used as a  testing theoretical laboratory.

The {\it Toy Model} above  has four active flavours: the three light quarks $u$, $d$, $s$ and the heavy charm quark with mass $M_c$. On the other hand, the Marichev interpolations that we have discussed in the previous section are for three light flavours $u$, $d$, $s$. Therefore, in order to compare it with the  {\it Toy Model}, we have to subtract from it the charm quark $c$ contribution. This we do by observing that the effective  charm-quark contribution is well described by a constituent charm quark model with a spectral function
\be\lbl{eq:spcharm}
\frac{1}{\pi}\Imm\Pi_{c}(t)=\frac{\alpha}{\pi}\frac{1}{3}\ N_c \left(\frac{4}{9}\right) \left(1+\frac{2 M_c^2}{t}\right)\sqrt{1-\frac{4 M_c^2}{t}}\ \theta(t-4 M_c^2)\,,
\ee
with (we shall use for $M_c$  its central value)
\be
M_c = (1.275\pm 0.025)~\GeV\,.
\ee
This gives a contribution to the Mellin transform:

{\setl
\bea\lbl{eq:mellcharm}
\cM^{\rm charm}(s) & = & \int_{t_0}^\infty \frac{dt}{t}\left(\frac{t}{t_0}\right)^{s-1}\frac{1}{\pi}\Imm\Pi_{c}(t)
= \left( \frac{4 M_c^2}{t_0}\right)^{s-1}\int_{4 M_c^2}^\infty \frac{dt}{t}\left(\frac{t}{4 M_c^2}\right)^{s-1}\frac{1}{\pi}\Imm\Pi_{c}(t)
\nn \\
 & = & \frac{\alpha}{\pi}\frac{1}{3}\ \Nc\left( \frac{4}{9}\right)\left( \frac{4 M_c^2}{t_0}\right)^{s-1}\frac{1}{1-s}\frac{3\sqrt{\pi}}{4}\frac{\Gamma(3-s)}{\Gamma(\frac{7}{2}-s)}\,,\quad -\infty\le \Ree(s)<1\,,
\eea}

\noi
and the Mellin transform of the  {\it Toy Model} spectral function to compare with should, therefore, be:
\be
\cM(s)=\cM^{\rm ToyM}(s)-\cM^{\rm charm}(s)\,,
\ee
which results in  the following values for the constants $\cA$ and $\cR$ defined in Eqs.~\rf{eq:A} and \rf{eq:R}:
\be
\cA=0.449485\quad\annd\quad \cR=0.661645\,.
\ee
Furthermore, the contribution to the muon anomaly from the charm spectral function in Eq.~\rf{eq:spcharm} with $M_c =1.275~\GeV$, using e.g. the standard representation in Eq.~\rf{eq:standard},  is 
\be\lbl{eq:charmsub}
a_{\mu}^{\rm HVP} ({\rm charm})=0.1094352\times 10^{-8}\,,
\ee
which also has to be subtracted as well from the one 
in Eq.~\rf{eq:amutoy}. More precisely,  the predictions for the muon anomaly using the  Marichev interpolation will be compared to the value
\be
a_{\mu}^{\rm HVP}=a_{\mu}^{\rm HVP}({\rm ToyM}) -a_{\mu}^{\rm HVP}({\rm charm})=6.70274\times 10^{-8}\,.
\ee

At this level it is interesting to compare the Mellin transform of the {\it Toy Model} with those corresponding to the two Marichev interpolations in Eqs.~\rf{eq:M120} and \rf{eq:M120m1}. This is shown in Fig.~\rf{fig:mellinTMV} below. One can see a net improvement from the first interpolation (the green curve) to the second one (the blue curve) which is already quite close to the {\it Toy Model} one (the red curve).

%%%%%%%%%%%%%%%%%%%%%%%%%%%%%
\begin{figure}[!ht]
\begin{center}
\hspace*{-1cm}\includegraphics[width=0.60\textwidth]{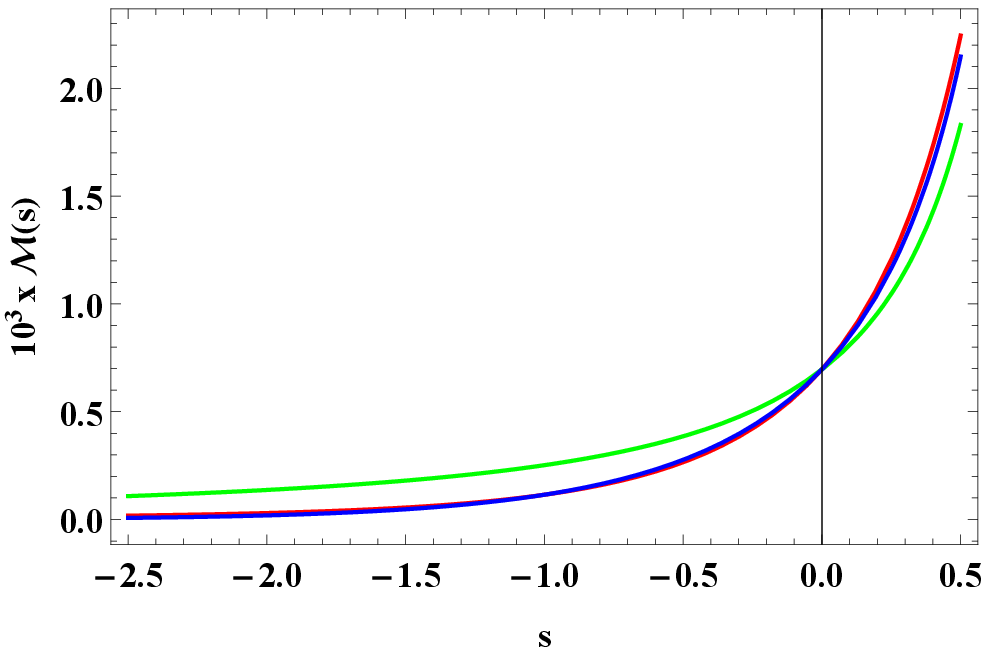} 
\bf\caption{\lbl{fig:mellinTMV}}
\vspace*{0.25cm}
{\it Mellin Transforms of the Hadronic Spectral Function\\
Red: Toy Model without Charm \\  Green:   Marichev's Interpolation with only $\cM(0)$ as input\\ Blue:   Marichev's Interpolation with $\cM(0)$ and $\cM(-1)$ as input. }
\end{center}
\end{figure}
%%%%%%%%%%%%%%%%%%%%%%%%%%%%%%%
%%%%%%%%%%%%%%%%%%%%%%%%%%%%%
\begin{figure}[!ht]
\begin{center}
\hspace*{-1cm}\includegraphics[width=0.60\textwidth]{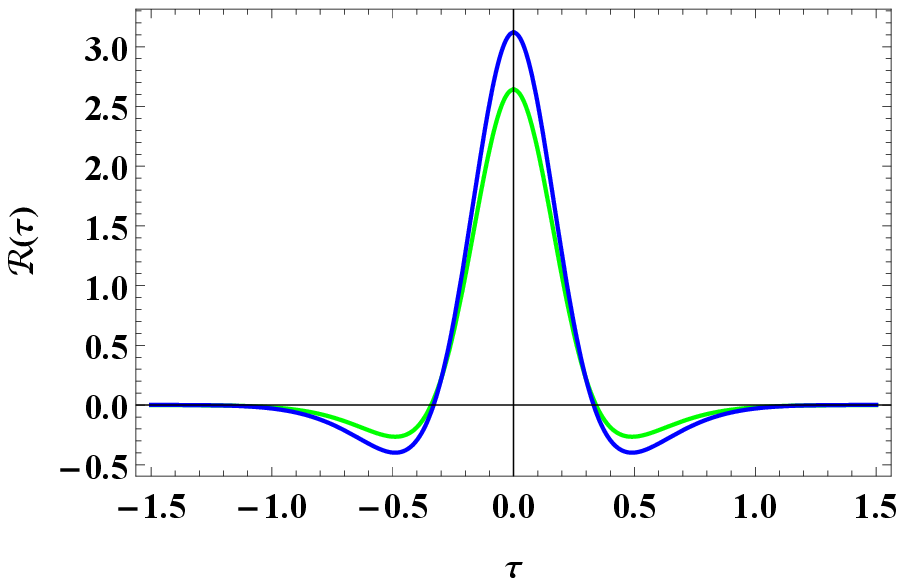} 
\bf\caption{\lbl{fig:integrands12}}
\vspace*{0.25cm}
{\it Shape of the Functions $\Re^{1,2}_{0}(\tau)$ (green curve)and  $\Re^{1,2}_{0,-1}(\tau)$ (blue curve)}
\end{center}
\end{figure}
%%%%%%%%%%%%%%%%%%%%%%%%%%%%

It is also interesting to show the shapes of the real parts of the integrands in Eqs.~\rf{eq:mb1mar} and \rf{eq:mb2mar} as  functions of $\tau$ corresponding to the {\it First} and {\it Second Marichev Interpolations}, i.e. the functions

{\setl
\bea
\Re^{1,2}_{0}(\tau) & = & \frac{1}{\cC}\left(\frac{m_{\mu}^2}{t_0} \right)^{-\left(\frac{1}{2}-i\tau \right)} \cF\left(\frac{1}{2}-i\tau \right)\ \cM^{(1,2)}_{(0)}\left(\frac{1}{2}-i\tau \right)\lbl{eq:intfuntau1}\\
\Re^{1,2}_{0,-1}(\tau) & = & \frac{1}{\cC}\left(\frac{m_{\mu}^2}{t_0} \right)^{-\left(\frac{1}{2}-i\tau \right)} \cF\left(\frac{1}{2}-i\tau \right)\ \cM^{(1,2)}_{(0,-1)}\left(\frac{1}{2}-i\tau \right)\,, \lbl{eq:intfuntau2}
\eea}

\noi
where for convenience we have factorized the overall constant $\cC$ in Eq.~\rf{eq:constantc}. The shapes of the real parts of these two functions of $\tau$ are shown in Fig.\rf{fig:integrands12}. The green curve corresponds to $\Re^{1,2}_{0}(\tau)$, the blue curve to $\Re^{1,2}_{0,-1}(\tau)$

The result we get for $a_{\mu}^{\rm HVP}$ using the {\it First Marichev Interpolation} given by Eq.~\rf{eq:M120}, i.e. the one   corresponding to the curves in green in Figs.~\rf{fig:mellinTMV} and \rf{fig:integrands12}, is
\be
a_{\mu}^{\rm HVP}({\rm first})=6.25021\times 10^{-8}\,.
\ee
It reproduces the {\it Toy Model} value at the 6.6\% level. Not competitive enough for a comparison with the experimental results, but a net improvement with respect to the upper bound~\cite{BdeR69} value:
\be
a_{\mu}^{\rm HVP} \le \left(\frac{\alpha}{\pi}\right) \frac{m_{\mu}^2}{t_0} \frac{1}{3}\cM(0)= 7.72132\times 10^{-8} \,.
\ee
Notice that, at this level of approximation, i.e. with only $\cM(0)$ known, there is no possible prediction from Pad\'e approximants. 

Things get much better at the level of the {\it Second Marichev Interpolation} which results in the value
\be
a_{\mu}^{\rm HVP}({\rm second})=6.74591\times 10^{-8}\,,
\ee  
and reproduces the {\it Toy Model} value at the 0.6\% level. We find this very encouraging!

\subsection{\normalsize Test with the BHLS Model of ref.~\cite{BDDJ16}.}

\noi
One may perhaps suspect that the reason for the success of the previous results is due to the particular choice of  the {\it Toy Model}  as a reference. We have, therefore,  also considered another phenomenological model as an alternative reference: the so called BHLS-Model of ref.~\cite{BDDJ16}, and applied the same method to it. For that we choose the entries corresponding to what the authors of ref.~\cite{BDDJ16} call {\it Data Direct}. The values quoted for the first two moments are:

{\setl 
\bea
\cM(0)_{\rm BHLS} & = & (10.1307\pm 0.0745)\times 10^{-5}\,,\\
\cM(-1)_{\rm BHLS} & = & (0.23507\pm 0.00185)\times 10^{-5}\,,
\eea}

\noi
and the corresponding result for the muon anomaly which they find is
\be\lbl{eq:BHLSan}
a_{\mu}^{\rm HVP}({\rm BHLS})=(683.50\pm 4.75)\times 10^{-10}\,.
\ee
The central values of the BHLS-moments, in our normalization ($t_0=4 m_{\pi^{\pm}}^2$), correspond to
\be
\cM(0)  =  0.707094\times 10^{-3} \quad\annd\quad 
\cM(-1) = 0.011452\times 10^{-3}\,.
\ee

Using these moments as an input, the result for $a_{\mu}^{\rm HVP}$ from the first Marichev approximation, the one which only requires $\cM(0)$ as an input, is
\be
a_{\mu}^{\rm HVP}({\rm first})=626.12\times 10^{-10}\,.
\ee
Using the second Marichev approximation, which requires both $\cM(0)$ and $\cM(-1)$ as input, we find
\be
a_{\mu}^{\rm HVP}({\rm second})=676.32\times 10^{-10}\,,
\ee
in agreement with the BHLS value in Eq.~\rf{eq:BHLSan} at the 1\% level. With the charm contribution in Eq.~\rf{eq:charmsub}
subtracted to the BHLS value in Eq. (6.18) the agreement is at the 0.6\% level, much the same
as in the case of the Toy Model.

\subsection{\normalsize An Application to the LQCD Prediction of ref.~\cite{Lellouch16}.}

\noi
Let us now apply the Marichev interpolation technique to recent LQCD results. 
The lattice QCD BMWc collaboration has recently published results on the first two moments $\cM(0)$ and $\cM(-1)$. Their numbers~\cite{Lellouch16}:
\be
\Pi_{1}[\GeV^{-2}]=0.0999(10)(9)(23)(13)\quad\annd\quad
\Pi_{2}[\GeV^{-4}]=-0.181(6)(4)(10)(2)\,,
\ee
when expressed in the normalization ($t_0=4 m_{\pi^{\pm}}^2$) of our Mellin Moments, with the charm contribution subtracted,  and with their errors added quadratically, correspond to the values:
\be\lbl{eq:latmoments}
\cM(0)_{{\cite{Lellouch16}}}=(0.704\pm 0.021)\times 10^{-3}\quad\annd\quad
\cM(-1)_{{\cite{Lellouch16}}}=(0.101\pm 0.007)\times 10^{-3}\,.
\ee
These numbers result in the following values for the parameters 
\be\lbl{eq:AR}
\cA=0.455\pm 0.014 \quad\annd\quad \cR= 0.572\pm 0.043\,,
\ee
in Eqs.~\rf{eq:A} and \rf{eq:R} and are the ones 
to be inserted in the {\it First} and {\it Second  Marichev Interpolations} in Eqs.~\rf{eq:M120} and \rf{eq:M120m1}.
%%%%%%%%%%%%%%%%%%%%%%%%%%%%%
\begin{figure}[!ht]
\begin{center}
\hspace*{-1cm}\includegraphics[width=0.60\textwidth]{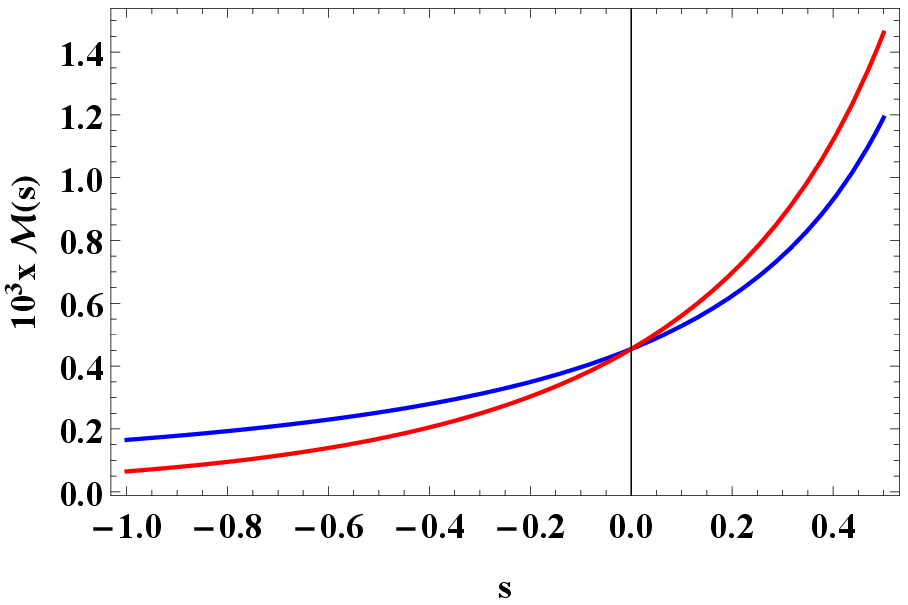} 
\bf\caption{\lbl{fig:MTL}}
\vspace*{0.25cm}
{\it Mellin Transforms corresponding to the Marichev Interpolations\\   using as an input the central values of the  LQCD results of ref.~\cite{Lellouch16}.\\
Blue is the first interpolation,  Red the second one.}
\end{center}
\end{figure}
%%%%%%%%%%%%%%%%%%%%%%%%%%%%%%%

Figure~\rf{fig:MTL} shows the Mellin transforms obtained with the central values of the numbers above; here the blue curve corresponds to the {\it First-Interpolation}, the red curve to the improved 
{\it Second-Interpolation}. These curves are rather similar to the ones in Fig.~\rf{fig:mellinTMV} which test the {\it Toy Model}.
%%%%%%%%%%%%%%%%%%%%%%%%%%%%%
\begin{figure}[!ht]
\begin{center}
\hspace*{-1cm}\includegraphics[width=0.60\textwidth]{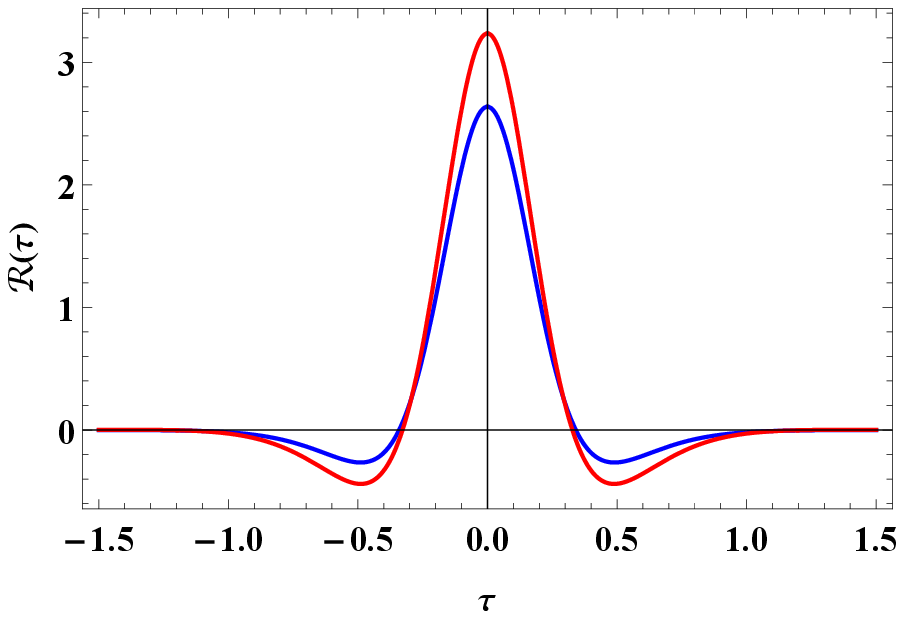} 
\bf\caption{\lbl{fig:intsl}}
\vspace*{0.25cm}
{\it Shape of the Functions $\Re^{1,2}_{0}(\tau)$ (blue curve)and  $\Re^{1,2}_{-1,0}(\tau)$ (red curve)\\   using as an input the central values of the LQCD results of ref.~\cite{Lellouch16}.}
\end{center}
\end{figure}
%%%%%%%%%%%%%%%%%%%%%%%%%%%%

Figure~\rf{fig:intsl} shows the shapes of the real parts of the functions $\Re^{1,2}_{0}(\tau)$ and $\Re^{1,2}_{0,-1}(\tau)$ in Eqs.~\rf{eq:intfuntau1} and \rf{eq:intfuntau2} using the central values of the LQCD results in ref.~\cite{Lellouch16}. It is interesting that, although these shapes differ in detail from the ones corresponding to the {\it Toy Model} in Fig.~\rf{fig:integrands12} and from the QED one in Fig.~\rf{fig:tauqed} they are {\it qualitatively} rather similar. This is probably due to the fact that they all have in common the Gamma-function structure of the Marichev general ansatz in Eq.~\rf{eq:marichev}.

 The corresponding predictions for $a_{\mu}^{\rm HVP}$ using the {\it first} and {\it second} interpolations, with the values of the moments in Eq.~\rf{eq:latmoments} in which the charm contributions have been subtracted, are:
\be
a_{\mu}^{\rm HVP}({\it First}) = (6.23\pm 0.18)\times 10^{-8} \,,
\ee
and
\be
a_{\mu}^{\rm HVP}({\it Second)} = (6.81\pm 0.30 )\times 10^{-8} \,.
\ee
The error in $a_{\mu}^{\rm HVP}({\it First})$ is an average of the two limits of error induced by the error in $\cM(0)$. The error in $a_{\mu}^{\rm HVP}({\it Second)}$ is sensitive to both the errors in $\cM(0)$ and $\cM(-1)$; it  has been estimated by evaluating $a_{\mu}^{\rm HVP}$ with the limit of errors in $\cM(0)$ in Eq.~\rf{eq:latmoments} keeping the central value of $\cM(-1)$, then evaluating $a_{\mu}^{\rm HVP}$ with the limit of errors in $\cM(-1)$ in Eq.~\rf{eq:latmoments} keeping the central value of $\cM(0)$ and finally averaging the partial errors quadratically. We find these results very encouraging to pursue with more accurate LQCD determinations of $\cM(0)$ and $\cM(-1)$ and, if possible, with the determination of higher moments.

%%%%%%%%%%%%%%%%%%%%%%%%%%%%%%%%%%
\section{\large Conclusions.}
\setcounter{equation}{0}
\def\theequation{\arabic{section}.\arabic{equation}}

\noi
We conclude from this work  that with a precise  determination of the first Mellin Moment $\cM(0)$ i.e., with a precise determination of just the slope of the HVP function at the origin accessible to LQCD, one can already obtain a result for $a_{\mu}^{\rm HVP}$ which provides a first rough test of the determinations using experimental data. Notice that in the {\it First Marichev Interpolation}, besides the eventual  determination of $\cM(0)$, one only uses as further information two well known properties of QCD: asymptotic freedom and the fact that in the chiral limit there is no $1/Q^2$ term in the OPE of $\Pi(Q^2)$. With such limited input there is no prediction from Pad\'e approximants one can compare with. 

The {\it Second Marichev Interpolation} of the Mellin Transform of the hadronic spectral function which we have developed above  results in a much more accurate determination. It includes as an input the determinations of  the first two moments $\cM(0)$ and $\cM(-1)$, i.e. the determination of the first two derivatives of the HVP function $\Pi(Q^2)$ at $Q^2 =0$ accessible to LQCD. The test with the  {\it Toy Model} above results in a determination  of $a_{\mu}^{\rm HVP}$ with an accuracy of $0.6\%$ which is very encouraging. The application to the  determination of the $\cM(0)$ and $\cM(-1)$ moments from LQCD~\cite{Lellouch16} points towards a very promising future in this direction.

It would be very helpful to be able to test the {\it Marichev Interpolation Approach}  with real experimental data. In that respect we encourage our colleagues of refs.~\cite{Davier11,Hagiwara11} to publish the values of a few moments: $\cM(0)$, $\cM(-1)$, $\cM(-2)\,,\cdots$ of the same physical spectral function which they use for their determination of $a_{\mu}^{\rm HVP}$.   

%\vspace*{1.2cm} 
\newpage

\begin{center}
{\normalsize\bf Acknowledgments.}
\end{center}
\vspace*{0.25cm}

I am very much indebted to David Greynat for many helpful comments and suggestions on the topics discussed in this paper; in particular for bringing to my attention Ramanujan's Master Theorem.
I have also benefited from  discussions with Laurent Lellouch, who has kindly provided the code of his {\it Toy Model},  and with Marc Knecht and Santi Peris. I wish to thank David Greynat and Laurent Lellouch for a careful reading of the successive versions of the manuscript.

\vspace*{1.2cm}

%%%%%%%%%%%%%%%%%%%%%%%%%%%%%%%%%%%%%%%%


\begin{thebibliography}{99}

\bibitem{LPdeR72}
         B.E.~Lautrup, A.~Peterman and E.~de Rafael, Phys. Rep. {\bf C3} (1972) 193.
								
\bibitem{EdeR94}
         E.~de Rafael, Phys. Lett. {\bf B322} (1994) 239.
				
\bibitem{KPdeR98}
         M.~Knecht, S.~Peris and E.~de Rafael, Phys. Lett. 
				 {\bf B443} (1998) 255.                                 

\bibitem{Lowetal67}
         T.~Das, G.S.~Guralnik, V.S.~Mathur, F.E.~Low and J.E.~         Young,  Phys. Rev. Lett. {\bf 18} (1967) 759. 
								
\bibitem{EdeR03}
         E.~de Rafael, Nucl. Phys. (Proc. Suppl.) 
				 {\bf B119} (2003) 71.
\bibitem{BNL}	
         G.W.~Bennett et al. (The $g$-2 Collab.), Phys. ReV. {\bf D73} (2006) 072003.
				
\bibitem{TH}
         Th.~Blum, A.~Denig, I.~Logashenko, E.~de Rafael, B.~Lee Roberts, Th.~Teubner and G.~Venanzoni, {\it The Muon $(g-2)$ Theory Value: Present and Future}, arXiv:1311.2198v1 [hep-ph].				
																
\bibitem{Davier11}
         M.~Davier, A.~Hoecker, B.~Malaescu, and Z.~Zhang, 
				  Eur. Phys.J. {\bf C71} (2011) 1515.	
							
\bibitem{Hagiwara11}
        K.~Hagiwara, R.~Liao, A.D.~Martin, D.~Nomura and T.K.~Teubner, J. Phys {\bf G38} (2011) 085003.
									
\bibitem{Davier16}
				M.~Davier, {\it Update of the Hadronic Vacuum Polarization Contribution to the muon $g-2$}, arXiv:1612.02743v2 [hep-ph].

\bibitem{Burger14}
         F.~Burger, X.~Feng, G.~Hotzel, K.~Jansen, M.~Petschlies and D.B.~Renner, (ETM Collaboration), JHEP {\bf 02} (2014) 099.	
																										
\bibitem{Ch16}
         B.~Chakraborty, C.T.H.~Davis, P.G.~de Oliveira, J.~         Koponen and G.P.~Lepage, (HPQCD collaboration), arXiv:1601.03071 [hep-lat].	
												
\bibitem{Lellouch16}
         Sz.~Borsanyi, Z.~Fodor, T.~Kawanai, S.~Krieg, L.~Lellouch, R.~Malak, K.~Miura, K.K.~Szabo, C.~Torrero and B.~Toth, arXiv:1612.02364v1 [hep-lat].
				
			
				
\bibitem{ToyM}
         L.~Lellouch, {\it Private Comunication}.						
													
\bibitem{Raman}
        B. Berndt. {\it Ramanujan's Notebooks, Part I}. Springer-Verlag, New York, 1985.
																		
																		
\bibitem{Mari82}
        O.I.~Marichev, {\it Handbook of Integral Transforms  of Higher Transcendental Functions: Theory and Algorithmic Tables}, Ellis Horwood Ltd., Chichester, U.K. 198.																			

\bibitem{Poul10}
         {\it Transforms and Applications Handbook}, Third Edition,  Editor-in-Chief, Alexander D.~Poularikas, CRC Press 2010. Ch~12: J~Bertrand, P.~Bertrand and J-Ph~Ovarlez, {\it Mellin Transform}.
         
\bibitem{Slater66}
         L.J.~Slater, {\it Generalized Hypergeometric Functions}, Cambridge University Press, 1966.
					
\bibitem{GdeR17}
        D.~Greynat and E.~de Rafael, {\it The Mellin-Barnes Approach to Hadronic Vacuum Polarization and $g_{\mu}-2$}, ( to be published).																	
				
\bibitem{EdeR14}
         E.~de Rafael, Phys. Letters {\bf B736} (2014) 52.		
						
							
\bibitem{BDDJ16}
         M.~Benayoun, P.~David, L.~DelBuono and F.~Jegerlehner, 
				 arXiv:1605.04474v1 [hep-ph].							
																	
\bibitem{FGD95}
         Ph.~Flajolet, X.~Gourdon and Ph.~Dumas, Theor. Comput., Sci. 
				 {\bf144} (1995) 3.
					
\bibitem{FGdeR05}
         S.~Friot, D.~Greynat, and E.~de Rafael, Phys. Lett. {\bf B68} (2005) 73. 	
						
\bibitem{AGdeR08}
         J.Ph.~Aguilar, D.~Greynat and E.~de Rafael, Phys. Rev. {\bf D77} (2008) 093010.			
																	
\bibitem{FG12}
         S.~Friot and D.~Greynat, J.~Math. Phys. {\bf 53} (2012) 
				 023508.
				
\bibitem{BM61}
         C.~Bouchiat and L.~Michel, J.~Phys.Radium, {\bf 22} (1961) 121.	
							
\bibitem{BdeR69} 
         J.S. Bell and E. de Rafael, Nucl. Phys. {\bf B11} 611 (1969).
								
\bibitem{Blum03}
         T.~Blum, Phys. Rev. Lett. {\bf 91} (2003) 052001.
								
								
\bibitem{Perisetal12}
         C.~Aubin, T.~Blum, M.~Golterman and S.~Peris, Phys. Rev. {\bf D86} (2012) 054509.
								
\bibitem{GMP14}
        M.~Golterman, K.~Maltman and S.~Peris, Phys. Rev. {\bf D90} (2014) 074508.
												
\bibitem{ABCGPT16}
        C.~Aubin,T.~Blum, P.~Chau, M.~Golterman, S.~Peris and C.~Tu, Phys. Rev. {\bf        D93} (2016) 05450. 
											
                                    	                      

\end{thebibliography}
\end{document}